\documentclass[12pt]{iopart}

\usepackage{amsfonts,amstext,iopams,cancel,xcolor}

\newcommand{\Lie}{\mathcal{L}}

\begin{document}

\title{Coordinate independent expression for transverse trace-free tensors}



\author{Rory Conboye}

\address{Department of Physics, Florida Atlantic University, Boca Raton,
FL 33431-0991}

\begin{abstract}
The transverse and trace-free (TT) part of the extrinsic curvature represents half of the dynamical degrees of freedom of the gravitational field in the $3 + 1$ formalism. As such, it is part of the freely specifiable initial data for numerical relativity. Though TT tensors in 3-space possess only two component degrees of freedom, they cannot ordinarily be given solely by two scalar potentials. Such expressions have been derived however, in coordinate form, for all TT tensors in flat space which are also translationally or axially symmetric (Conboye and \'O Murchadha 2014 \emph{Class. Quantum Grav.} {\bf 31} 085019). Since TT tensors are conformally covariant, these also give TT tensors in conformally-flat space. In this article, the work above has been extended by giving a coordinate-\emph{independent} expression for these TT tensors. The translational and axial symmetry conditions have also been generalized to invariance along any hypersurface orthogonal Killing vector.

\

\noindent{Keywords: transverse traceless tensors, initial-value problem, gravitational degrees of freedom}
\end{abstract}

\pacs{04.20.Ex, 04.20.Cv, 02.40.Hw}

\section{Introduction}

On a 3-dimensional Riemannian manifold, Helmholtz' theorem gives an orthogonal decomposition of a vector field into its curl-free and divergence-free parts, known respectively as the longitudinal and transverse parts of the vector field. With appropriate conditions, the curl-free part can be given by the divergence of a scalar field, showing the longitudinal part to have one component degree of freedom. The divergence-free condition on the transverse part then constrains one of it's three components, leaving two component freedoms.

Symmetric 2-index tensors can also be decomposed, this time into three parts, as shown by Deser \cite{Deser67,Deser68} and York \cite{YorkTT}. A tensor can first be separated into its trace and trace-free parts. The latter is then decomposed into a tensor with the same divergence as the original trace-free part, and a divergence-free tensor known as transverse and trace-free (TT),
\begin{equation}\label{TTDecomp}
T^{a b}
 = T g^{a b} + T_L^{a b} + T_{TT}^{a b} ,
\end{equation}
\setcounter{equation}{0}
such that
\numparts
\begin{eqnarray}
 &T = g_{a b} T^{a b} ,
 \qquad
 &\nabla_a T_L^{a b} = \nabla_a (T^{a b} - g^{a b} T) , \\
 &g_{a b} T_{TT}^{a b} = 0 ,
 \qquad
 &\nabla_b T_{TT}^{a b} = 0 . \label{TTCond}
\end{eqnarray}
\endnumparts
This decomposition is orthogonal if the 3-manifold on which the tensor is defined is simply-connected, or compact without boundary. For a TT tensor, the vector-valued divergence-free and scalar-valued trace-free conditions \eref{TTCond} give 4 constraints on the 6 independent components of a symmetric 2-index tensor. A TT tensor can therefore be expected to have 2 component freedoms, the same as a transverse vector field.

In the $3 + 1$ formalism of General Relativity (GR), the space-time manifold is foliated into space-like hypersurfaces. The kinematics of each hypersurface is given by the spatial 3-metric $g_{a b}$ and its extrinsic curvature $K^{a b}$. The later gives the embedding of the hypersurface in the space-time 4-manifold, and is directly related to the conjugate momentum of the spatial metric. Decomposing the Einstein equations gives a set of constraint equations for each hypersurface, given in a vacuum as
\numparts
\begin{eqnarray}
R + K^2 - K_{a b} K^{a b} = 0 , \label{Hamiltonian} \\
\nabla_b (K^{a b} - g^{a b} K) = 0 . \label{momentum}
\end{eqnarray}
\endnumparts
These are known as the Hamiltonian and momentum constraints, with $R$ representing the 3-scalar curvature and $K = g_{a b} K^{a b}$ the trace of the extrinsic curvature. These constraints restrict $4$ of the combined $12$ components (both $g_{a b}$ and $K^{a b}$ being symmetric 2-index tensors on a 3-manifold). With $4$ components essentially related to the choice of coordinates, this leaves only $4$ \emph{true} component freedoms for the gravitational field. These can also be interpreted as $2$ \emph{dynamical} degrees of freedom.

In the canonical form of vacuum electromagnetism, the constraint parts of Maxwell's equations amount to setting the longitudinal parts of the electric field and vector potential to zero. The dynamical degrees of freedom are then given by their transverse parts, which are canonically conjugate to one another. In section 4 of Arnowit, Deser and Misner's seminal work \cite{ADM}, the TT tensor decomposition is used to give a similar canonical form for GR. However, since the metric is itself a dynamical field in GR, the divergence and trace must be taken with respect to some background metric. The TT parts of $g_{a b}$ and $K^{a b}$ are then shown to give the dynamical degrees of freedom of the gravitational field, as a deviation from this background metric. In general, if the trace of the curvature is zero (known as maximal slicing), the momentum constraint \eref{momentum} reduces to a divergence-free condition for the curvature $\nabla_b K^{a b} = 0$, with the divergence now taken with respect to the spatial metric $g_{a b}$. In this case $K^{a b}$ is both transverse and trace-free with respect to $g_{a b}$ itself, without reference to any background metric.

TT tensors are also conformally covariant, so that for any positive function $\phi$,
\begin{eqnarray}
&T^{a b} \qquad &\text{TT \ w.r.t} \qquad g_{a b} \nonumber \\
\Leftrightarrow \qquad
&\bar{T}^{a b} = \phi^{-10} T^{a b} \qquad &\text{TT \ w.r.t.} \qquad
 \bar{g}_{a b} = \phi^4 g_{a b} . \label{ConformalCov}
\end{eqnarray}
This property was used by York et al. \cite{York71,OMYork,YorkTT} to develop the Conformal Transverse Trace-free (CTT) decomposition, separating free and constrained initial data in $3 + 1$ GR, in a background-independent manner. The TT part of the conformal extrinsic curvature represents part of the freely specifiable initial data, with the conformal factor and longitudinal part of the curvature solved for using the constraint equations (2). The Bowen-York curvature \cite{BY} is a specific solution of the CTT decomposition, which is commonly used for initial data in numerical relativity.

Though TT tensors on a 3-manifold contain only two component freedoms, there is no method in general for obtaining them from a choice of two scalar potentials alone. The standard method for the CTT decomposition involves choosing an initial symmetric 2-index tensor $M^{a b}$ (6 components) and solving an adapted momentum constraint so that the divergence of $M^{a b}$ is subtracted from the longitudinal part of the curvature (for more details, see for example sections 3.2 of both text books \cite{Alc,BS}). Since 6 components need to be chosen, where there should only be 2 free choices, changes to even a number of these components could easily lead to the same initial conditions as before. In contrast, any expression depending on 2 scalar potentials alone, gives \emph{direct} control over the component freedoms of the initial data, and hence the relevant physical degrees of freedom of the system.

Progress in this direction has been made by Dain \cite{Dain}, where the existence of a surface-orthogonal Killing vector is assumed, and a TT tensor expression depending on a \emph{single} scalar potential is given. This expression represents all TT tensors invariant along this Killing vector, as long as an additional space-time symmetry of the Killing vector is also assumed. This time-symmetry, outlined in \cite{BrandtSeidel}, essentially removes the second component freedom. A more general class of solutions to the momentum constraint \eref{momentum} is considered in \cite{TafelJoz14,Tafel15} using tensors which are transverse, though not necessarily trace-free, but reduce to the expression in \cite{Dain} for the same conditions.

In earlier work involving the author \cite{Potentials}, coordinate expressions depending on \emph{two} scalar potentials are derived. These are shown to represent \emph{all} TT tensors which are either translationally or axially symmetric in flat or conformally flat space. One of these potentials is equivalent to that of \cite{Dain}, with the second potential replacing the time-symmetry restriction required there. The axially-symmetric coordinate expression derived in \cite{Potentials} also appears in \cite{KM04}, where it is shown to be a special case of a large class of asymptotically flat initial data proved to exist in \cite{DF01}.

In this article, a coordinate-\emph{independent} form is found for the coordinate expressions derived in \cite{Potentials}. The symmetry conditions are also generalized to tensors which are unchanged along any surface-orthogonal Killing vector, for which translational and axial symmetries are special cases. Even though the expression is only TT in flat space, a corresponding TT tensor for any \emph{conformally}-flat metric can always be given using \eref{ConformalCov}. Conformally-flat initial data has proved particularly beneficial in numerical relativity due to its superposition properties \cite{BrandtBrug}, with the tensor expression in this article giving direct control over the momentum degrees of freedom for conformally-flat systems with the appropriate symmetries.


In section \ref{sec:Foliations} the symmetry conditions are outlined and the equations for a surface-orthogonal Killing vector derived. Section \ref{sec:earlier} gives the relevant results from both \cite{Dain} and \cite{Potentials}, with the later used to motivate the proposed TT tensor expression in section \ref{sec:expression}, which is constructed to be trace-free. A simplified form of the tensor is derived in section \ref{sec:div}, and its divergence shown to vanish for a flat metric, proving the proposed tensor to be TT. The main result can be found immediately from \eref{TT}, excluding the part given in \cite{Dain}. A complete and simplified form of the TT tensor, depending on two scalar potentials, is then given by \eref{TT 2}.

\section{Foliating Symmetries}
\label{sec:Foliations}

The symmetry conditions considered in this article involve a foliation of the spatial 3-manifold into isometric hypersurfaces. For these hypersurfaces to be isometric, there must exist a Killing vector $\eta^a$ along which the metric remains unchanged, i.e. the Lie derivative with respect to $\eta^a$, denoted $\Lie_{\vec \eta}$, is zero:
\begin{equation}\label{Killing}
\Lie_{\vec \eta} \, g_{a b} \
 = \ \nabla_a \eta_b + \nabla_b \eta_a \
 =  \ 0 \ ,
\end{equation}
which is known as Killing's equation. A vector or tensor is considered symmetric along $\eta^a$ if its Lie derivative with respect to $\eta^a$ is zero. It can easily be shown that the Lie derivatives of the vector $\eta^a$ itself, its corresponding covector $\eta_a$ and the norm $\eta := g_{a b} \eta^a \eta^b$, must all vanish along $\eta^a$. It can also be shown that the second covariant derivative of a Killing vector is related to the Riemann curvature tensor,
\begin{equation}\label{Killing-Riemann}
\nabla_a \nabla_b \, \eta_c \
 = \ R_{c b a}^{\ \ \ d} \, \eta_d \ .
\end{equation}
The Leibniz product rule for the covariant derivative also gives the relation
\begin{equation}\label{grad eta}
\nabla_a \eta \
 = \ 2 \, \eta^b \nabla_a \eta_b \
 = \ - 2 \, \eta^b \nabla_b \eta_a \ ,
\end{equation}
with the second part coming from \eref{Killing}. These relations will be used in section \ref{sec:div}.

A Killing field foliates a manifold into isometric hypersurfaces if and only if Frobenius' theorem is satisfied (see for example \cite{KobayNomizuII}, or Appendix B of \cite{Wald}). In this case, there must exist some 1-form $\btheta$ such that $d\bfeta = \bfeta \wedge \btheta$, with $\bfeta$ seen here as a 1-form, and $d$ representing the exterior derivative. In terms of covariant derivatives and abstract index notation, there must exist a 1-form $\theta_a$ such that
\begin{equation}\label{Frobenius}
\nabla_a \eta_b - \nabla_b \eta_a \
 = \ \eta_b \, \theta_a - \eta_a \, \theta_b \ .
\end{equation}
Since $\eta^a$ is a Killing vector, the left-hand side can be reduced using \eref{Killing},
\begin{equation}\label{Frobenius2}
2 \, \nabla_a \eta_b \
 = \ \eta_b \, \theta_a - \eta_a \, \theta_b \ .
\end{equation}
To find an expression for $\theta_a$, first observe that any component of $\theta_a$ parallel to $\eta_a$ must vanish on the right-hand side of the equation. It can therefore be assumed that $\theta_a$ is orthogonal to $\eta_a$ so that $\eta^a \theta_a = 0$. Applying the vector $\eta^b$ to both sides,
\begin{equation}
2 \, \eta^b \nabla_a \eta_b \
 = \ \eta \, \theta_a - \eta_a \, \cancelto{0}{\eta^b \theta_b} \ ,
\end{equation}
and using \eref{grad eta} on the left-hand side, the expression $\theta_a = \frac{1}{\eta} \, \nabla_a \eta$ is found. This is a necessary form for $\theta_a$, if it exists such that \eref{Frobenius2} is satisfied. Substituting this into \eref{Frobenius2},
\begin{equation}\label{Foliation}
2 \, \eta \, \nabla_a \eta_b \
 =  \ \eta_b \nabla_a \eta
    - \eta_a \nabla_b \eta \ ,
\end{equation}
a manifold is foliated into isometric hypersurfaces orthogonal to $\eta^a$ if and only if this equation is satisfied. This equation can also be found in \cite{Wald} (Appendix C (C.3.12)), where its form and derivation differ slightly from here.

\section{Earlier results}
\label{sec:earlier}

The tensor expressions from \cite{Dain} and \cite{Potentials} are given explicitly in this section, to aid referencing and the consistency of the notation.

\subsection{Time-Killing symmetry}
\label{sec:Dain}

A TT tensor depending on a single scalar potential is given by Dain in \cite{Dain} by reformulating work from \cite{BakerPuzio} in a coordinate-independent form. This tensor is symmetric along a 2-surface orthogonal Killing vector, as outlined above. The tensor also satisfies a `time - Killing' \emph{space-time} symmetry $(t,\eta^a) \rightarrow (-t, -\eta^a)$ when it represents the extrinsic curvature of a space-like hypersurface, as discussed in \cite{BrandtSeidel}.

The extrinsic curvature tensor, from (A1) of \cite{Dain}, is given as
\begin{equation}\label{Dain K}
K^{a b} \
 =  \ \frac{1}{\eta}
      \left(S^a \eta^b + S^b \eta^a\right) \ ,
\end{equation}
for a space-like hypersurface with a spatial vector $\eta^a$ satisfying \eref{Killing} and \eref{Foliation}. This tensor is both transverse and trace-free if the vector field $S^a$ is subject to the equations
\begin{equation}\label{Dain S constraints}
\Lie_{\vec \eta} \, S^a \
 = \ 0 \, , \quad
S^a \eta_a \
 = \ 0 \, , \quad
\nabla_a S^a \
 = \ 0 \, .
\end{equation}
A solution for $S^a$ is given by (A6) of \cite{Dain}, in terms of a scalar potential $\omega$,
\begin{equation}\label{Dain S solution}
S^a \
 =  \ \frac{1}{\eta} \, \epsilon^{a b c}
      \eta_b \, \nabla_c \omega \ ,
	\qquad
      \Lie_{\vec \eta} \, \omega \
 =  \ 0 \ ,
\end{equation}
where $\epsilon_{a b c}$ is the Levi-Civita tensor, the unit alternating tensor times $\sqrt{g}$, where $g$ represents the determinant of the spatial $3$-metric. Combining \eref{Dain K} and \eref{Dain S solution} gives the only non-zero components of the extrinsic curvature for a space-time with a time-rotation symmetry, as shown by equation (16) of \cite{BrandtSeidel}.

\subsection{Coordinate expressions with two potentials}
\label{sec:pot}

Expressions in specific coordinate systems, depending on two scalar potentials alone, were derived in an earlier work \cite{Potentials}. These expressions were shown to give all TT tensors, which are either translationally or axially symmetric, on a flat simply-connected or compact 3-manifold. For convenience, these expressions are given below in matrix form, with some minor notational adjustments from \cite{Potentials}.

In Cartesian coordinates $(x, y, z)$ the TT equations \eref{TTCond} and the translational symmetry condition $\Lie_{\vec z} \, T^{a b} = 0$ (with $\Lie_{\vec z}$ representing the Lie derivative with respect to the z-coordinate vector field) are used to derive the following ((17) from \cite{Potentials}),
\begin{equation}\label{TT L Car}
T^{a b} \ = \
\left(
  \begin{array}{ccc}
        \partial_{y y} \, \nu
     &- \partial_{x y} \, \nu
     &- \partial_y \, \omega \\
   & & \\
      - \partial_{x y} \, \nu
     &  \partial_{x x} \, \nu
     &  \partial_x \, \omega \\
   & & \\
      - \partial_y \, \omega
     &  \partial_x \, \omega
     &- \partial_{x x} \, \nu - \partial_{y y} \, \nu \\
  \end{array}
 \right) \ .
\end{equation}
This expression is TT with respect to a flat metric, and symmetric along the $z$-coordinate as long as the scalar potentials $\omega$ and $\nu$ are invariant along $z$, that is $\Lie_{\vec z} \, \omega = \Lie_{\vec z} \, \nu = 0$. A coordinate transformation then leads to an expression in cylindrical-polar coordinates ($\rho, \phi, z)$ ((30) from \cite{Potentials}),
\begin{equation}\label{TT L Cyl}
\fl T^{a b} =
\left(
  \begin{array}{ccc}
        \frac{1}{\rho^2} \ \partial_{\phi \phi} \, \nu
       + \frac{1}{\rho} \ \partial_\rho \, \nu
     &- \frac{1}{\rho^2} \ \partial_{\rho \phi} \, \nu
       + \frac{1}{\rho^3} \ \partial_\phi \, \nu
     &- \frac{1}{\rho} \ \partial_\phi \, \omega \\
   & & \\
      - \frac{1}{\rho^2} \ \partial_{\rho \phi} \, \nu
       + \frac{1}{\rho^3} \ \partial_\phi \, \nu
     &  \frac{1}{\rho^2} \ \partial_{\rho \rho} \, \nu
     &  \frac{1}{\rho} \ \partial_\rho \, \omega \\
   & & \\
      - \frac{1}{\rho} \ \partial_\phi \, \omega
     &  \frac{1}{\rho} \ \partial_\rho \, \omega
     &- \partial_{\rho \rho} \, \nu
       - \frac{1}{\rho} \ \partial_\rho \, \nu
       - \frac{1}{\rho^2} \ \partial_{\phi \phi} \, \nu \\
  \end{array}
 \right) \ .
\end{equation}

In cylindrical $(\rho, z, \phi)$ and spherical-polar coordinates $(r, \theta, \phi)$ with an axial symmetry condition $\Lie_{\vec \phi} \, T^{a b} = 0$, (52) and (53) from \cite{Potentials} give the expressions
\begin{equation}\label{TT A Cyl}
\fl T^{a b} =
\left(
  \begin{array}{ccc}
         \frac{1}{\rho^2} \ \partial_{z z} \, \nu
       - \frac{1}{\rho^3} \ \partial_\rho \, \nu
     & - \frac{1}{\rho^2} \ \partial_{\rho z} \, \nu
     &   \frac{1}{\rho^3} \ \partial_z \, \omega \\
   & & \\
       - \frac{1}{\rho^2} \ \partial_{\rho z} \, \nu
     &   \frac{1}{\rho^2} \ \partial_{\rho \rho} \, \nu
       - \frac{1}{\rho^3} \ \partial_\rho \, \nu
     & - \frac{1}{\rho^3} \ \partial_\rho \, \omega \\
   & & \\
         \frac{1}{\rho^3} \ \partial_z \, \omega
     & - \frac{1}{\rho^3} \ \partial_\rho \, \omega
     & - \frac{1}{\rho^4} \ \partial_{\rho \rho} \, \nu
       - \frac{1}{\rho^4} \ \partial_{z z} \, \nu
       + \frac{2}{\rho^5} \ \partial_{\rho} \, \nu \\
  \end{array}
 \right) \ ,
\end{equation}
\begin{equation}\label{TT A Sph}
\fl T^{a b} =
\left(
  \begin{array}{ccc}
      \frac{1}{r^4 \sin^2 \theta} \ \partial_{\theta \theta} \, \nu
  & - \frac{1}{r^4 \sin^2 \theta} \ \partial_{r \theta} \, \nu
  & - \frac{1}{r^4 \sin^3 \theta} \ \partial_\theta \, \omega
		\\
    - \frac{\cos \theta}{r^4 \sin^3 \theta} \ \partial_\theta \, \nu
  & + \frac{1}{r^5 \sin^2 \theta} \ \partial_\theta \, \nu
  & 		\\
 & & 		\\
    - \frac{1}{r^4 \sin^2 \theta} \ \partial_{r \theta} \, \nu
  &   \frac{1}{r^4 \sin^2 \theta} \ \partial_{r r} \, \nu
    - \frac{1}{r^5 \sin^2 \theta} \ \partial_r \, \nu
  &   \frac{1}{r^4 \sin^3 \theta} \ \partial_r \, \omega
		\\
    + \frac{1}{r^5 \sin^2 \theta} \ \partial_\theta \, \nu
  & - \frac{\cos \theta}{r^6 \sin^3 \theta} \ \partial_\theta \, \nu
  &  		\\
 & & 		\\
    - \frac{1}{r^4 \sin^3 \theta} \ \partial_\theta \, \omega
  &   \frac{1}{r^4 \sin^3 \theta} \ \partial_r \, \omega
  & - \frac{1}{r^4 \sin^4 \theta} \ \partial_{r r} \, \nu
    + \frac{1}{r^5 \sin^4 \theta} \ \partial_r \, \nu
		\\
  &
  & - \frac{1}{r^6 \sin^4 \theta} \ \partial_{\theta \theta} \, \nu
    + \frac{2 \cos \theta}{r^6 \sin^5 \theta} \ \partial_\theta \, \nu
		\\
  \end{array}
 \right) ,
\end{equation}
with $\omega$ and $\nu$ independent of the coordinate $\phi$, so that $\Lie_{\vec \phi} \, \omega = \Lie_{\vec \phi} \, \nu = 0$. Note that the potential $\nu$ is related to the potential $R_A$ used in \cite{Potentials} by the equation $\nu := - \rho R_A = - r \sin \theta R_A$.

The TT expressions in \cite{KM04} are given by choosing $\nu = W / r$ and $\omega = Z$, in spherical coordinates \eref{TT A Sph}. A relation to the results of \cite{DF01} can also be found in \cite{KM04}. The components depending on the potential $\omega$ in the matrix expressions above can be given by \eref{Dain K} and \eref{Dain S solution}, i.e. the expression from \cite{Dain}. The remaining components, those depending on $\nu$, vanish in the expression given in \cite{Dain} due to the space-time symmetry condition in \cite{BrandtSeidel}.

\section{Coordinate independent expression}
\label{sec:expression}

In this section, the results of \cite{Dain} and \cite{Potentials} displayed in section \ref{sec:earlier} above are used to motivate a coordinate-independent expression which is hypothesized to be TT. This proposed TT tensor is constructed to be trace-free, and its divergence is then computed in the next section and shown to vanish in flat space, proving the hypothesized expression to be TT.

\subsection{Matching coordinate expressions}

In the tensor expressions derived in \cite{Potentials}, the coordinates are chosen so that the Killing vector coincides with one of the coordinate vectors, $z$ in \eref{TT L Car}, \eref{TT L Cyl} and $\phi$ in \eref{TT A Cyl}, \eref{TT A Sph}. The resulting expressions contain terms involving the potential $\omega$ only in the off-diagonal $z / \phi$ components. Terms involving $\nu$ are then found in the non-$z / \phi$ components (upper-left $2 \times 2$ parts), and in the $z z / \phi \phi$ component.

To find a coordinate independent expression for the components involving $\nu$, similar to \eref{Dain K} for those involving $\omega$, the different expressions in section \ref{sec:pot} are analysed for patterns. From this, the double derivatives in the components of each tensor are found to be given by the expression
\begin{equation}
A^{a b} \
 = \ \epsilon^{a i k} \epsilon^{b j l} \, \eta_i \eta_j \,
	\partial_k \partial_l \, \nu \ ,
\end{equation}
with the contractions of the Levi-Civita symbols with the Killing covectors ensuring that the Killing vector components vanish. It seems natural to normalize this expression by factoring out the norm of the Killing covectors, and to use a covariant instead of ordinary derivative, leading to
\begin{equation}
A^{a b} \
 = \ \frac{1}{\eta^2} \epsilon^{a i k} \epsilon^{b j l} \eta_i \eta_j
	\nabla_k \nabla_l \, \nu \ .
\end{equation}
This expression exactly matches the components of the translational symmetry tensors \eref{TT L Car} and \eref{TT L Cyl}, and matches the double derivative terms for the axially-symmetric tensors.

For the diagonal non-Killing components of \eref{TT A Cyl} and \eref{TT A Sph}, it was observed that the remaining terms are equivalent to the connection coefficients of the derivative $\nabla_\phi \nabla_\phi \nu$, with a similar normalization term of $\frac{1}{\eta^2}$. Conveniently, this term vanishes for a translational symmetry, with the new expression
\begin{equation}\label{Aab}
A^{a b} \
 = \ \frac{1}{\eta^2} \left(
	  \epsilon^{a i k} \epsilon^{b j l} \eta_i \eta_j
	- g^{a b} \eta^k \eta^l
	\right) \nabla_k \nabla_l \, \nu \ ,
\end{equation}
giving the non-$z / \phi$ components for \emph{all} of the tensors in section \ref{sec:pot}.

\subsection{Ensuring a trace-free tensor expression}

For a symmetry-adapted coordinate system, any remaining terms in the final $\eta \eta$ component (bottom-right component for expressions in section \ref{sec:pot}), can be given by the tensor
\begin{equation}
B^{a b} \ = \ \eta^a \eta^b \, \beta \ .
\end{equation}
This tensor must be defined so that the tensor $T^{a b} = A^{a b} + B^{a b}$ is trace-free,
\begin{equation}
0 \
 = \ g_{a b} T^{a b} \
 = \ g_{a b} A^{a b} + g_{a b} \, \eta^a \eta^b \, \beta \
 = \ g_{a b} A^{a b} + \eta \, \beta \ .
\end{equation}
This implies that the scalar function $\beta$ must be defined as
\begin{equation}
\beta \,  \
 := \ - \frac{1}{\eta} \, g_{a b} A^{a b} \
  = \ - \frac{1}{\eta^3} \left(
	  g_{a b} \epsilon^{a i k} \epsilon^{b j l} \eta_i \eta_j
	- 3 \, \eta^k \eta^l
	\right) \nabla_k \nabla_l \, \nu \ ,
\end{equation}
which gives a trace-free tensor expression for $T^{a b}$.

A complete coordinate-independent trace-free tensor expression is now given by
\begin{eqnarray}\label{TT}
T^{a b} \
 &=  \ \frac{1}{\eta^2} \left(
	  \epsilon^{a i k} \epsilon^{b j l} \eta_i \eta_j
   	- g^{a b} \eta^k \eta^l \right)
	\nabla_k \nabla_l \, \nu \nonumber \\
 & \ - \frac{1}{\eta^3} \eta^a \eta^b \left(
	  g_{c d} \epsilon^{c i k} \epsilon^{d j l} \eta_i \eta_j
	- 3 \, \eta^k \eta^l \right)
	\nabla_k \nabla_l \, \nu \ ,
\end{eqnarray}
for any scalar function $\nu$ with $\Lie_{\vec \eta} \, \nu = 0$. For a vanishing potential $\omega$, the coordinate expressions in section \ref{sec:pot} are given completely by this expression. This tensor will be shown in the next section to be divergence-free in flat-space, showing it to be TT.

\section{Finding the divergence of the expression}
\label{sec:div}

In order to show that the expression above is transverse, and that it satisfies the Killing symmetry of section \ref{sec:Foliations}, the expression is first simplified. Both the divergence and the Lie derivative with respect to $\eta^a$ are then taken, and shown to vanish for a flat metric.

To begin, the relationship between the Levi-Civita symbol and the metric are required. The product of two Levi-Civita symbols can be given as
\begin{eqnarray}\label{epsilon pair}
\epsilon^{a i k} \epsilon^{b j l} \
  &= \ g^{a b} g^{i j} g^{k l} - g^{a b} g^{i l} g^{k j}
	\nonumber \\
 & \ + g^{a j} g^{i l} g^{k b} - g^{a j} g^{i b} g^{k l}
	\nonumber \\
 & \ + g^{a l} g^{i b} g^{k j} - g^{a l} g^{i j} g^{k b} \ .
\end{eqnarray}
A contracted product, appearing in the second line of \eref{TT}, reduces even further,
\begin{equation}\label{epsilon pair contraction}
g_{c d} \epsilon^{c i k} \epsilon^{d j l} \
  = \ g^{i j} g^{k l} - g^{i l} g^{k j} \ ,
\end{equation}
by directly applying $g_{c d}$ to \eref{epsilon pair}. Second order derivatives of the scalar potential can also be reduced to first order derivatives, when contracted with the Killing vector,
\begin{equation}\label{eta eta deriv}
\eta^k \eta^l \, \nabla_k \nabla_l \, \nu \
 = \ \eta^k \nabla_k \,
      \cancelto{0}{\left(\eta^l \nabla_l \, \nu\right)} \
  - \ \eta^k \, \nabla_k \eta^l \, \nabla_l \, \nu \
 = \ \frac{1}{2} \, \nabla^l \eta \, \nabla_l \, \nu \ ,
\end{equation}
with the first part given by the product rule and the cancelling due to $\Lie_{\vec \eta} \, \nu = 0$, and the second coming from the use of \eref{grad eta}. Another contraction can be shown to vanish,
\begin{eqnarray}\label{eta grad deriv}
\eta^k \nabla^l \eta \ \nabla_k \nabla_l \, \nu \
 &= \ - \frac{1}{2 \eta} \nabla^l \eta \, \nabla_l \eta \
	\cancelto{0}{\eta^k \nabla_k \, \nu} \ \
    + \frac{1}{2 \eta} \, \cancelto{0}{\eta_l \nabla^l \eta} \
	\nabla^k \eta \, \nabla_k \, \nu \
  = \ 0 \ ,
\end{eqnarray}
by application of \eref{Foliation}, with the cancelling due to $\Lie_{\vec \eta} \, \nu = \Lie_{\vec \eta} \, \eta = 0$. This expression also vanishes if $\eta$ and $\nu$ are swapped.

\subsection{Simplifying tensor expression}

The metric expressions for the Levi-Civita symbols are first substituted into \eref{TT},
\begin{eqnarray}
T^{a b} \
 &= 
   \ - \frac{1}{\eta} \nabla^a \nabla^b \nu
   \ + \frac{1}{\eta} g^{a b} \nabla^2 \nu
   \ - 2 \frac{1}{\eta^2} \eta^a \eta^b \nabla^2 \nu
	   \nonumber \\ &
   \ + \frac{1}{\eta^2} \eta^a \eta^k \nabla^b \nabla_k \nu
   \ + \frac{1}{\eta^2} \eta^b \eta^k \nabla^a \nabla_k \nu
	   \nonumber \\ &
   \ - 2 \frac{1}{\eta^2} g^{a b} \eta^k \eta^l \nabla_k \nabla_l \nu
   \ + 4 \frac{1}{\eta^3} \eta^a \eta^b \eta^k \eta^l \nabla_k \nabla_l \nu
 \ .
\end{eqnarray}
The chain rule and $\Lie_{\vec \eta} \, \nu = \eta^a \nabla_a \nu = 0$ are used to reduce the second covariant derivatives of $\nu$ in the second line, with \eref{Foliation} then used to reduce derivatives of $\eta^a$ to derivatives of $\eta$. In the third line above, \eref{eta eta deriv} is used to reduce the second derivatives of $\nu$ to first derivatives, so that
\begin{eqnarray}\label{TT 1}
\fl T^{a b} \
 &= \ - \frac{1}{\eta} \nabla^a \nabla^b \nu
      + \left(g^{a b} - \frac{2}{\eta} \eta^a \eta^b \right)
	\frac{1}{\eta} \nabla^2 \nu
      - \left( g^{a b} - \frac{3}{\eta} \eta^a \eta^b \right) 
	\frac{1}{\eta^2} \nabla^k \eta \ \nabla_k \nu
	   \ ,
\end{eqnarray}
giving the final form of $T^{a b}$.

The tensor expression above can be seen as being generated by the tensor
\begin{equation}\label{Sab}
S^{a b} \
 = \   \frac{1}{\eta} \left( - \nabla^a \nabla^b \nu
     + g^{a b} \nabla^2 \nu
     - g^{a b} \frac{1}{\eta} \nabla^k \eta \nabla_k \nu \right) \ .
\end{equation}
This is first projected onto the 2-surfaces orthogonal to the Killing vector field $\eta^a$ using the projector $P^a_b = \delta^a_b - \eta^a \eta_b/\eta$. A trace-free tensor is then given by defining the part twice contracted with the Killing vector to give a vanishing trace for the full tensor
\begin{equation}
T^{a b} \
 = \ P^a_c P^b_d \ S^{c d} \
   - \ \frac{1}{\eta} \eta^a \eta^b \ g_{p q} (P^p_c P^q_d \ S^{c d}) \ .
\end{equation}
This form can easily be seen to agree with the tensors in section \ref{sec:pot} for appropriate choices of coordinates and the vector $\eta^a$.

\subsection{Divergence of tensor}

The divergence of \eref{TT 1} can now be computed. By applying the chain rule to each term, and cancelling using \eref{eta grad deriv} and $\Lie_{\vec \eta} \, \eta = \eta^a \nabla_a \eta = 0$, the divergence of $T^{a b}$ is given as
\begin{eqnarray}
\nabla_b T^{a b} \
 &= \
       \frac{1}{\eta} \left(
	 \nabla^a \nabla^b \nabla_b \nu - \nabla^b \nabla^a \nabla_b \nu
	\right)
	   \nonumber \\ & \
     - \frac{1}{\eta^2} \nabla^a \nabla^k \eta \ \nabla_k \nu
     + \frac{2}{\eta^3} \nabla^a \eta \ \nabla^k \eta \ \nabla_k \nu
	   \nonumber \\ & \
     - \frac{1}{\eta^2} \nabla^a \eta \ \nabla^2 \nu
     - \frac{2}{\eta^2} \eta^b \nabla_b \eta^a \ \nabla^2 \nu
	   \nonumber \\ & \
     + \frac{3}{\eta^3} \eta^b \nabla_b \eta^a \ \nabla^k \eta \ \nabla_k \nu
	\ . \label{divT 1}
\end{eqnarray}
The first line above is equivalent to $R^{a b \ k}_{\ \ b} \nabla_k \nu$, by the definition of the Riemann curvature tensor. Unfortunately this comes directly from the two second derivative terms of $S^{a b}$ \eref{Sab} which cannot be cancelled or removed. The curvature tensor must therefore be set to zero, restricting the 3-manifold to having a flat metric.

For the remaining terms in \eref{divT 1}, the covariant derivatives of the vector $\eta^a$ are reduced to derivatives of its norm $\eta$ using \eref{grad eta} and \eref{Foliation}, giving
\begin{equation}\label{divT 2}
\nabla_b T^{a b} \
 = \
     - \frac{1}{\eta^2} \nabla^a \nabla^k \eta \ \nabla_k \nu
     + \frac{1}{2 \eta^3} \nabla^a \eta \ \nabla^k \eta \ \nabla_k \nu
	\ .
\end{equation}
Taking part of the first term separately, and applying \eref{grad eta},
\begin{eqnarray}
\nabla^a \nabla^k \eta \ \nabla_k \nu \
 &= \ \nabla^a (2 \, \eta^l \nabla^k \eta_l) \ \nabla_k \nu
	\nonumber \\
 &= \ 2 \, \nabla^a \eta^l \ \nabla^k \eta_l \ \nabla_k \nu
    + 2 \, \eta^l \cancelto{0}{\nabla^a \nabla^k \eta_l} \ \nabla_k \nu
	\nonumber \\
 &= \ \frac{1}{2 \eta} \nabla^a \eta \ \nabla^k \eta \ \nabla_k \nu
	\ ,
\end{eqnarray}
with the cancelling in the second line due to \eref{Killing-Riemann} and the vanishing of the Riemann tensor. Substituting this back into \eref{divT 2},
\begin{equation}
\nabla_b T^{a b} \
 = \ - \frac{1}{2 \eta^3} \nabla^a \eta \ \nabla^k \eta \ \nabla_k \nu
     + \frac{1}{2 \eta^3} \nabla^a \eta \ \nabla^k \eta \ \nabla_k \nu \
 = \ 0 \ ,
\end{equation}
showing the divergence to vanish, for a flat metric, and proving the tensor $T^{a b}$ to be both transverse and trace-free.

The tensor expression can also be shown to be symmetric along the Killing vector $\eta^a$, by taking its Lie derivative with respect to $\eta^a$ and showing that it vanishes. This can be done quite easily, using a similar proceedure to that of the divergence, reducing all derivatives of $\eta^a$ and $\eta_a$ to derivatives of $\eta$ using \eref{grad eta} and \eref{Foliation}, and cancelling terms using \eref{eta eta deriv}, \eref{eta grad deriv} and the zero Riemann curvature tensor.



\section{Conclusion}

A coordinate-independent trace-free tensor expression has been given \eref{TT}, \eref{TT 1} and shown to be divergence-free on a flat 3-manifold, and hence TT. Combining this with the tensor expression given in \cite{Dain} (see \eref{Dain K} and \eref{Dain S solution} from section \ref{sec:Dain}) gives the coordinate-independent flat-space TT tensor expression
\begin{eqnarray}\label{TT 2}
T^{a b}
 &= 
     - \frac{1}{\eta} \nabla^a \nabla^b \nu
     + \left(g^{a b} - \frac{2}{\eta} \eta^a \eta^b \right)
	\frac{1}{\eta} \nabla^2 \nu
     - \left( g^{a b} - \frac{3}{\eta} \eta^a \eta^b \right) 
	\frac{1}{\eta^2} \nabla^k \eta \ \nabla_k \nu
	\nonumber \\ & \
   + \ \frac{1}{\eta^2} \left(
	  \eta^a \epsilon^{b i k}
   	+ \eta^b \epsilon^{a i k}
	\right) \eta_i \nabla_k \, \omega
	\ ,
\end{eqnarray}
which depends on two scalar potentials $\nu$ and $\omega$ alone, and a surface-orthogonal Killing vector $\eta^a$, with $\Lie_{\vec \eta} \, \nu = \Lie_{\vec \eta} \, \omega = 0$. The tensor expression is also symmetric along $\eta^a$, which can be seen by taking its Lie derivative along $\eta^a$, and showing it to vanish.

For a Killing vector giving either translational or axial-symmetry, the tensor above is exactly equivalent to the coordinate expressions derived in \cite{Potentials}. The new expression generalizes these by giving a coordinate-independent expression, and by generalizing the symmetry conditions to a broader class. For the tensor given by Dain \cite{Dain}, while in flat space, the new expression can be seen as removing the time-Killing symmetry of \cite{BrandtSeidel}. The expression is also related to the Bowen-York curvature tensor \cite{BY}, where the two real number variables representing the angular and linear momenta $J$ and $P$ are generalized to the scalar \emph{fields} $\omega$ and $\nu$ respectively. The specific choices of the potentials giving the Bowen-York solution were derived in coordinate form in \cite{Potentials}. The new expression gives the freedom, for example, to perform perturbations of the curvature about the Bowen-York solution, while ensuring that the Einstein constraints remain satisfied exactly.

While the requirement for a flat spatial metric may seem restrictive, an appropriate conformal transformation of \eref{TT 2} will be TT with respect to a conformally related metric, according to \eref{ConformalCov}. Also, for a flat spatial metric, tensors defined by different potentials and symmetric along different Killing vector fields, can be added pointwise to form new TT tensors. This is due to the linearity of both the trace and divergence, and the complete set of Killing vectors in flat space. For example, the flat metric contains axial symmetries about \emph{all} axes simultaneously, compared with the single axis for a more general axially symmetric metric. This is used extensively in Numerical Relativity, where two separate Bowen-York curvature tensors can be added to give binary black hole initial conditions on conformally flat spatial manifolds. The new tensor could also be used for both the metric and extrinsic curvature, when these are viewed as deviations from a conformally flat space, as outlined by Arnowit, Deser and Misner \cite{ADM}.

\ack

This research was supported by Air Force Research Laboratory Grant $\#$ FA8750-15-2-0047. I would like to thank Niall \'O Murchadha for many helpful discussions, and both Wolfgang Tichy and Warner A Miller for reading the manuscript and for some very useful suggestions. I would also like to dedicate this paper to my father Tony Conboye, my most influential teacher and one of my closest friends. Your example will never be forgotten.

\section*{References}

\bibliography{TT}
\bibliographystyle{unsrt}

\end{document}